\documentclass[sigconf]{acmart}

\usepackage{booktabs} % For formal tables

\usepackage{balance}

\acmConference[GISE'20]{1st international workshop on Governance in Software Engineering}{Seoul, South Korea}
\begin{document}
\title{Illuminating a Blind Spot in Digitalization -- Software Development in Sweden's Private and Public Sector}

\author{Markus Borg}
\orcid{XXX}
\affiliation{%
  \institution{RISE Research Institutes of Sweden}
  \streetaddress{P.O. Box 1212}
  \city{Lund}
  \country{Sweden}
  \postcode{43017-6221}
}
\email{markus.borg@ri.se}
\author{Joakim Wernberg}
\orcid{XXX}
\affiliation{%
  \institution{Swedish Entrepreneurship Forum}
  \streetaddress{P.O. Box 1212}
  \city{Stockholm}
  \country{Sweden}
  \postcode{43017-6221}
}
\email{joakim.wernberg@entreprenorskapsforum.se}
\author{Thomas Olsson}
\orcid{XXX}
\affiliation{%
  \institution{RISE Research Institutes of Sweden}
  \streetaddress{P.O. Box 1212}
  \city{Lund}
  \country{Sweden}
  \postcode{43017-6221}
}
\email{thomas.olsson@ri.se}
\author{Ulrik Franke}
\orcid{XXX}
\affiliation{%
  \institution{RISE Research Institutes of Sweden}
  \streetaddress{P.O. Box 1212}
  \city{Kista}
  \country{Sweden}
  \postcode{43017-6221}
}
\email{ulrik.franke@ri.se}
\author{Martin Andersson}
\orcid{XXX}
\affiliation{%
  \institution{Blekinge Institute of Technology}
  \streetaddress{P.O. Box 1212}
  \city{Karlskrona}
  \country{Sweden}
  \postcode{43017-6221}
}
\email{martin.andersson@bth.se}

% The default list of authors is too long for headers.
\renewcommand{\shortauthors}{M. Borg \textit{et al.}}

\begin{abstract}
As Netscape co-founder Marc Andreessen famously remarked in 2011, software is eating the world -- becoming a pervasive invisible critical infrastructure. Data on the distribution of software use and development in society is scarce, but we compile results from two novel surveys to provide a fuller picture of the role software plays in the public and private sectors in Sweden, respectively. Three out of ten Swedish firms, across industry sectors,  develop software in-house. The corresponding figure for Sweden's government agencies is four out of ten, i.e., the public sector should not be underestimated. The digitalization of society will continue, thus the demand for software developers will further increase. Many private firms report that the limited supply of software developers in Sweden is directly affecting their expansion plans. Based on our findings, we outline directions that need additional research to allow evidence-informed policy-making. We argue that such work should ideally be conducted by academic researchers and national statistics agencies in collaboration.
\end{abstract}

\copyrightyear{2020} 
\acmYear{2020} 
\setcopyright{acmcopyright}\acmConference[ICSEW'20]{IEEE/ACM 42nd International Conference on Software Engineering Workshops }{May 23--29, 2020}{Seoul, Republic of Korea}
\acmBooktitle{IEEE/ACM 42nd International Conference on Software Engineering Workshops (ICSEW'20), May 23--29, 2020, Seoul, Republic of Korea}
\acmPrice{15.00}
\acmDOI{10.1145/3387940.3392213}
\acmISBN{978-1-4503-7963-2/20/05}

%
% The code below should be generated by the tool at
% http://dl.acm.org/ccs.cfm
% Please copy and paste the code instead of the example below.
%



\maketitle

\section{Introduction}
Software is the key to leveraging the processing power and decentralized networks that characterize the digital infrastructure. It is what makes digital technology a so called general-purpose technology~\cite{bresnahan1995general} -- it enables an almost infinite variety of applications based on the same hardware. While software uptake and use pervade most of society and the economy, software development (SWD) still marks a frontline of digitalization. Organizations investing in SWD are signaling that they need more than off-the-shelf software products. 

SWD provides important indicators of digitalization and digital maturity that can be measured and compared across organizations. However, because such data is scarce, our understanding of digital transformation is left with significant blind spots. This is a considerable drawback to both researchers and policymakers, because a lack of data also inhibits the design and evaluation of policy measures aimed at promoting digitalization.

In this paper, we bring together two surveys aimed at gathering quantitative data on SWD in the private and the public sector, respectively. Two of the authors (Andersson and Wernberg) designed a novel survey that Swedsoft, a non-profit trade organization, commissioned from Statistics Sweden in 2017. The results have been summarized in a policy report by the two authors~\cite{andersson2018den}. The same year, the other authors (Borg, Olsson, and Franke) coincidentally conducted a census of SWD at Sweden's government agencies, also the first of its kind~\cite{borg2018digitalization}.

For the first time, we now combine the two studies to provide the most comprehensive perspective on SWD in Sweden to date. We summarize each survey and combine our findings to sketch further research questions -- particularly motivating an expansion of data collection of SWD in Sweden and elsewhere to enable evidence-informed policy-making~\cite{meissner2013science}. 

\section{Method}
\begin{figure}
\centering
\includegraphics[width=0.47\textwidth]{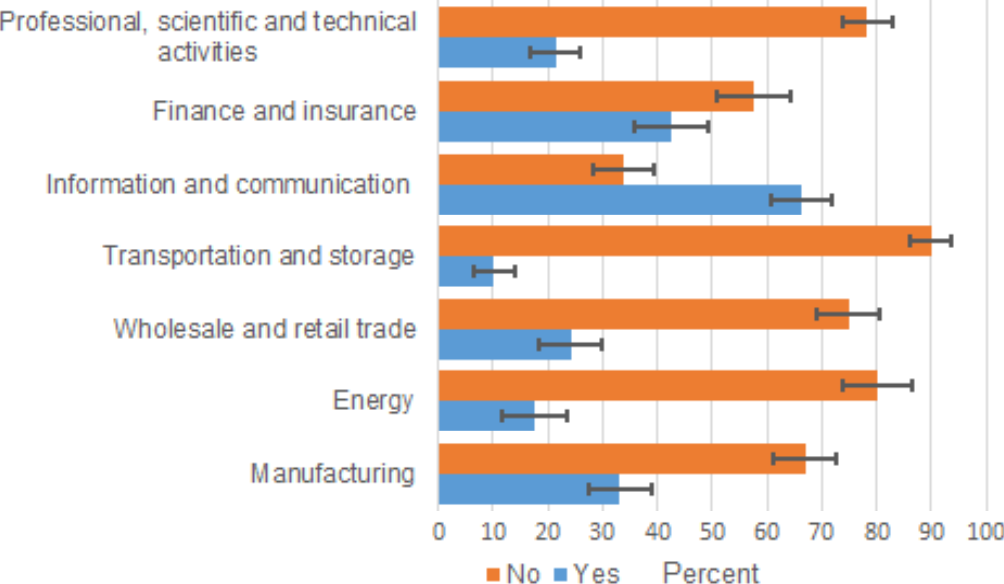}
\caption{In-house SWD in seven industry sectors.}
\label{fig:sectors}
\end{figure}

Statistics Sweden conducted the private sector survey. The scope of the survey covered SWD, recruitment and training of software developers, and the firms' expansion plans. A stratified sample of 3,000 firms was contacted out of the total population of 16,271 from 7 different industry sectors (cf.~\ref{fig:sectors}) with a response rate of 47\%. 
Statistics Sweden then scaled the responses to represent the entire population, meaning all results are associated with confidence intervals.

The public sector survey is based on a census conducted by requesting SWD documentation from 237 government agencies under the Freedom of the press act, 93 of which confirmed engaging in in-house SWD. We sent a questionnaire to study their development practices and 80 percent participated in the self-assessment, but some declined to provide answers due to national security concerns.

Combining the two studies gives a fuller picture of SWD in Sweden. Still, there are SWD activities in Sweden that are not covered. Examples include public sector development in municipalities, development in non-governmental organizations, and large non-profit open-source software projects.

\section{Software in Industry: Everywhere But Never the Same}
About 35\% of the surveyed firms report doing in-house SWD, either by bringing in contractors or having employed developers. SWD is spread across sectors, suggesting that digitalization is taking root across the economy -- but that it is unevenly picked up within sectors. We find the largest share of SWD in the ICT sector (65\%), followed by finance and insurance (40\%) and manufacturing (35\%).

Looking at firm size, 55\% of large firms in Sweden (>250 employees) develop software in-house, while the corresponding share among SMEs (10-49 or 50-249 employees) is about a third. These findings counterbalance the stereotypical intuition that large firms within the traditional manufacturing industries are not digital.

Firms developing software may also have something to show for it at the bottom line. Some research shows that increased software intensity is linked with more patents per R\&D dollar and a higher valuation on equity markets~\cite{branstetter2018get}. It would be worthwhile to study how digitalization moves from investment (cost) to affecting productivity across sectors and firm sizes -- we stress the need for more firm-level data and bottom line results.

\section{Identifying a Good Developer}
Another way of discerning differences across sectors is to look specifically at the demand for skills. It has become an oft-repeated mantra that there is a shortage of skilled people and that firms have a hard time attracting the relevant ``talent'', especially when it comes to engineers and computer scientists. Commonly discussed policy responses to address these issues include promoting engineering degrees more and matching university programs to the demand in the labor market. The survey results show, however, that this may prove easier said than done. 

\begin{figure}
\centering
\includegraphics[width=0.46\textwidth]{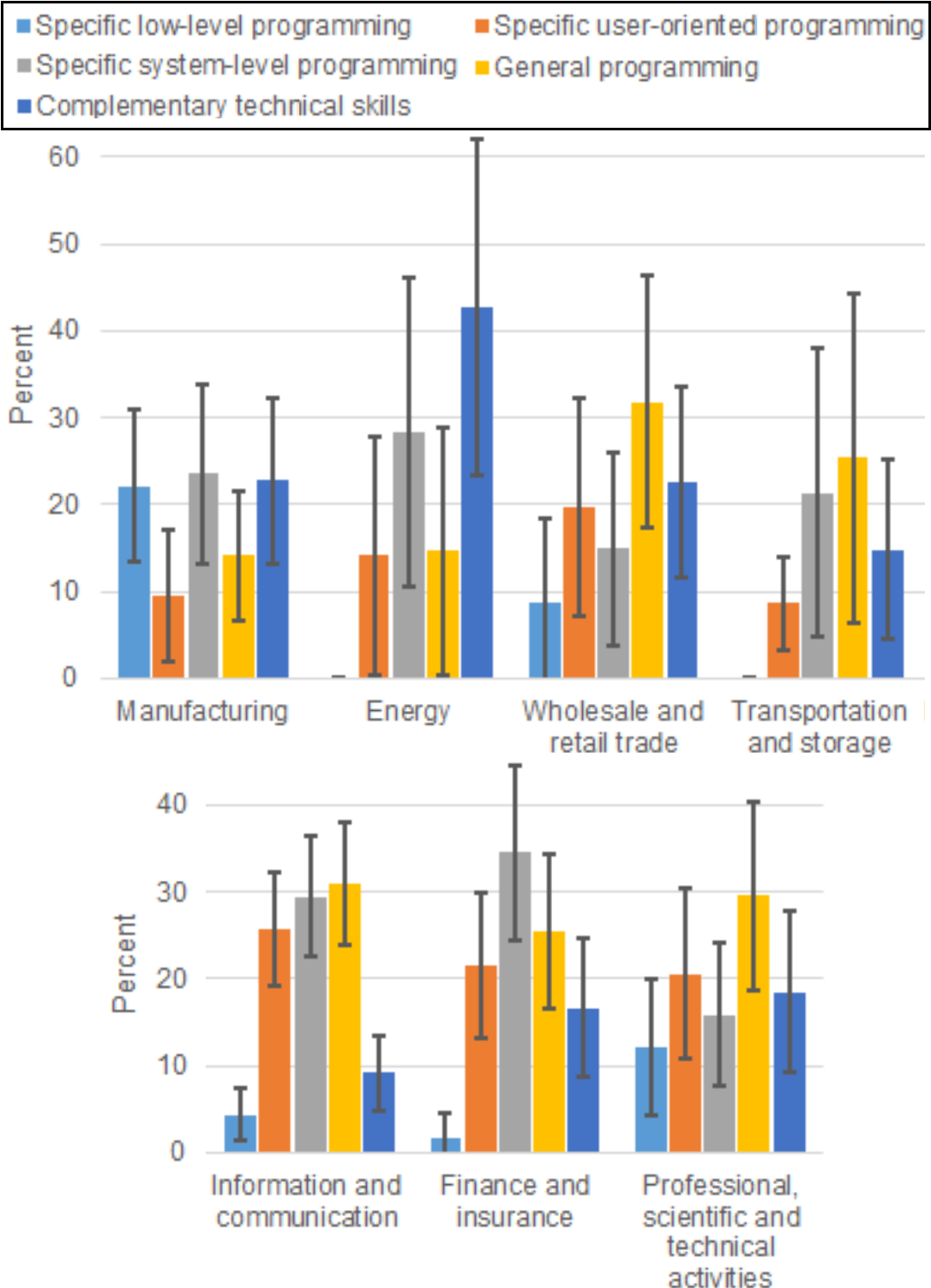}
\caption{Demanded SWD skills across sectors.}
\label{fig:skills}
\end{figure}

The demand for SWD skills is extremely heterogeneous across sectors (cf. Fig~\ref{fig:skills}), reflecting different types of SWD ranging from artificial intelligence and embedded software to app programming and e-commerce. 

Adding demand up across sectors, about 29\% prioritize general programming skills, followed by system-level programming and user-centered programming. While low-level programming was ranked low overall, it was highly valued among manufacturing firms -- a cornerstone in the Swedish economy. Again, we found extreme heterogeneity across, and within, sectors. Also, these results do not reflect the possible turn-over in specific types of skill demand. 
Looking at additional skills, firms' demands indicate a need for individuals who can orient themselves with respect to system architecture, user experience, and sector-specific knowledge. In terms of traits, firms prioritize analytical skills, creativity, and self-leadership -- all typical in programming jobs. However, they also seem to prioritize organizational skills and leadership, hinting at the need to work together on teams. 

Taking the numbers apart, the results suggest that firms are not just looking for any programmer but the right programmer. A similar pattern in high but indistinct demand for programming skills was documented in the US in the 1950s, following the computerization of industry~\cite{ensmenger2012computer}. It was then observed that while a good programmer can be extremely good, a poor programmer is disastrous, but it is hard to pin down the difference. Compared to the 1950s, SWD has moved from a niche occupation to becoming integrated within the entire economy ranging from mobile games to business models and critical infrastructure.

This goes to show that planning for the future demand of software-related skills is no straightforward task. A key challenge is to distinguish specific applications with short-term demand from long-term demand for basic skills. While the latter can be supplied by universities, the former may require other solutions. Gathering data on SWD and other related technologies across industries would open up several possible avenues of further research looking specifically at diffusion and change over time.

\section{Software in GovTech: Not to be Belittled}
The digital transformation impacts society at large. Just as firms must evolve to stay competitive on the market, also government agencies must adapt to meet novel digital expectations from their users. Agencies develop use technology, including software, to improve both internal efficiency and public services, aka. GovTech. Successful GovTech can boost transparency and citizen trust -- important concepts in democratic societies.

Thirty-nine percent of Sweden's government agencies develop software in-house. However, we interpret this figure as a lower bound, since 21 County Administrative Boards have centralized their development and several agencies rely on SWD in their respective parent agency. Furthermore, 16 agencies display considerable requirements engineering resources despite having no in-house SWD -- they develop advanced requirements specifications for public procurement.

The SWD organizations within the agencies are heterogeneous. While the most common number of developers is between 5 and 19, fourteen agencies report having more than 100 -- matching large firms. Also, the fraction of contractors in the development organizations varies across the full spectrum, but a majority have between 20-60\% contractors. Some agencies have none at all, others completely rely on contractors for SWD.

Looking at software sourcing strategies, the survey exhibits differences. Sixty-one percent of the agencies foci on in-house development to meet their software needs. At the same time, 58\% of the agencies report an ambition to purchase off-the-shelf solutions. Forty-two percent strive to integrate available open-source software into their solutions, but only 8\% develop software under open-source licenses -- a low number in the era of openness~\cite{runeson2019open}, and a possible target for further research.

\section{Best Practices in the Public Sector}
Government agencies are not stuck in old-fashioned development processes. Our study found evidence of popular approaches such as Scrum and Kanban. For example, 88\% of the agencies claim that their SWD is agile rather than plan-driven.

\begin{figure*}
\centering
\includegraphics[width=0.99\textwidth]{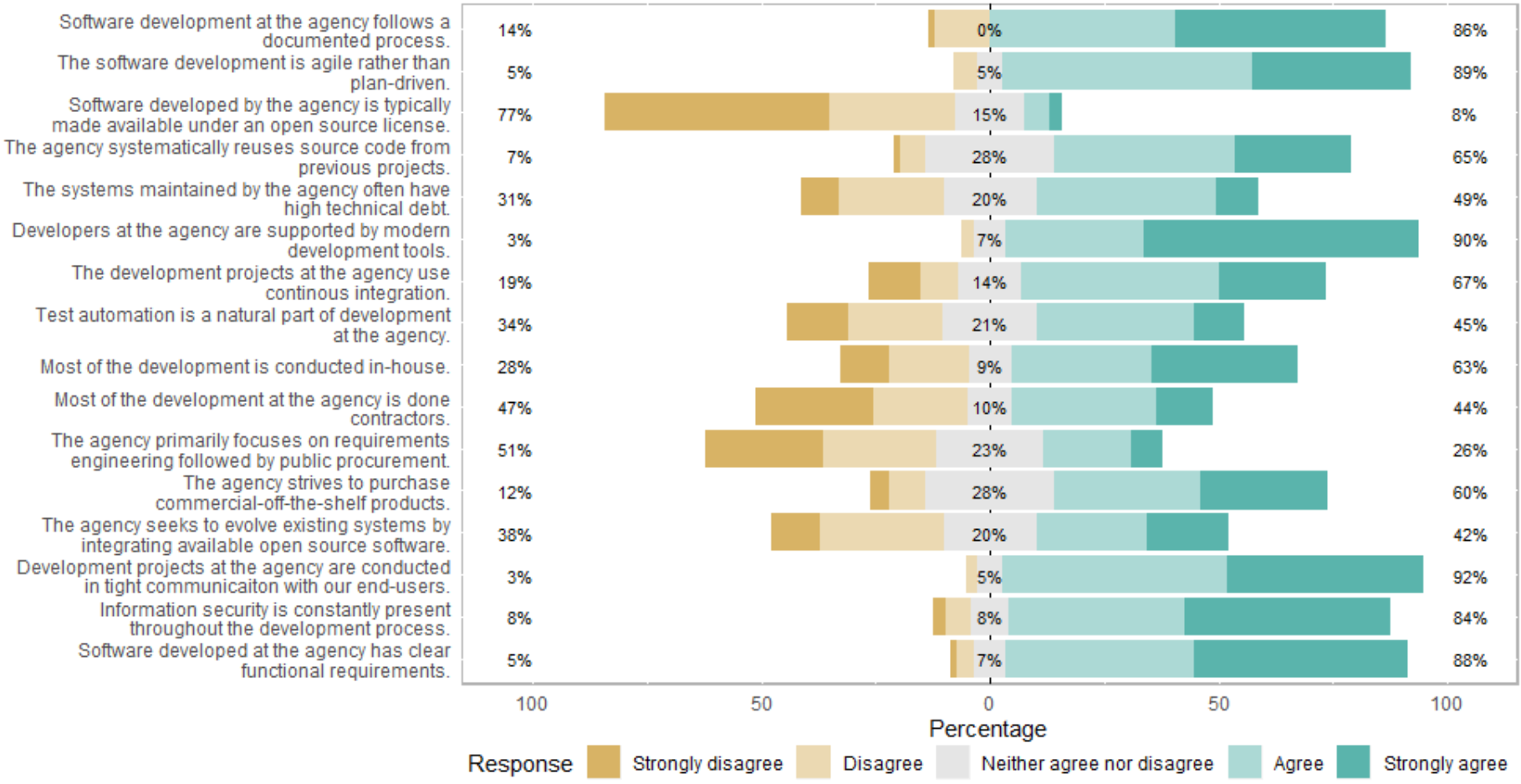}
\caption{Characteristics of SWD in the public sector. The bars show the level of agreement to statements.}
\label{fig:public}
\end{figure*}

Our results suggest that development in the public sector largely adheres to software engineering best practices. Fig.~\ref{fig:public} highlights results from our survey, and we note that a majority of agencies:

\begin{itemize}
\item Communicate frequently with the future end-users (91\%)
\item Develop sw guided by clear functional requirements (87\%)
\item Work with appropriate tool support (87\%)
\item Follow a documented development process (84\%)
\item Follow a security-aware development process (80\%)
\item Apply continuous integration (63\%) (but only 43\% rely on test automation)
\end{itemize}

Agencies need modern software engineering, as they struggle with complexity and scale just like firms. Our results demonstrate that public sector development faces some of the challenges that mostly have been documented in the private sector. Examples of ever-present software engineering challenging are legacies and short-termed technical solutions. Forty-seven percent of the agencies confirm the presence of high technical debt. We find a (weak) correlation between technical debt and in-house development primarily conducted by contractors -- another tempting target for future inquiry.

\section{One Snapshot of Digitalization to Show What We Don't Know}
Data on SWD and the spread of specific digital technologies is scarce, making it hard to study the progression of digitalization within and between different types of organizations. In this paper, we have brought together two novel surveys to provide the statistical starting point for better understanding what role SWD plays and will play in the Swedish economy and society today and in the future.

The combined results show that SWD, which marks an investment in the tailoring of digital tools as opposed to buying software off-the-shelf, occurs across industries and government agencies but is unevenly spread within sectors and departments. Interestingly, it appears to be within large firms and agencies rather than their smaller and presumably more agile peers that SWD is the most prevalent today.

Digitalization does not move forward one sector at a time. Differences between private and public organizations may be attributed to specific regulations and procedures related to public procurement. Nonetheless, it appears that digitalization is better described in terms of early adopters within sectors and departments than by leading and lagging sectors or leading private firms and lagging public agencies. 

The results imply highly diverse applications of SWD spanning both public departments with different focus and different industries. This is expected from a general-purpose technology like software. Furthermore, this suggests that the future demand for SWD skills will become more heterogeneous rather than more streamlined. The question is whether the educational policy should try to satisfy every specific niche demand for skills or instead attempt to identify long-term basic core skills of SWD. 

Against the backdrop of these two surveys, it is evident that the nature and dynamics of SWD, as well as the uptake of specific tools like artificial intelligence, marks a blind spot in much of our understanding of the digital transformation of the economy and society at large. It is not only about understanding the diffusion of a specific technology, but also about understanding the dynamics related to the various applications of a general-purpose technology across sectors and actors. There is a growing need for detailed, disaggregated data on how technologies are being utilized. 

\section{Conclusion} \label{sec:conc}
With increasing digitalization, the demand for software developers will escalate. Failing to meet this demand limits the growth of Sweden's economy, analogous to what has been identified as a root cause for the decline of the IT industry in Japan~\cite{arora2013going}. We propose six lines of further research that should encourage and motivate policymakers and national statistics agencies to collect data on the utilization of new technologies in the near future.

\begin{enumerate}
\item Identifying within-sector actor characteristics that drive early or late adoption of SWD and application of specific technologies. 
\item Will all organizations eventually engage in SWD?
\item The dynamical change in SWD across and within sectors over time. What characterizes sectors that adapt to utilizing new technologies faster?
\item What is the rate of change or turnover in specific (technical) skill demand? Is there a distinction between slow- and fast-changing skill demand with respect to utilizing new technologies?
\item How is work being reorganized with the utilization of, and adaptation to, new technologies?
\item What is the relation between digital maturity in public agencies and the corresponding private sector industries they monitor and work with? How has this relationship changed over time?
\end{enumerate}

\section*{Acknowledgement}
This study was partly funded by Swedsoft\footnote{www.swedsoft.se}, an independent non-profit organization with the mission to increase the competitiveness of Swedish software.

%\balance
\bibliographystyle{ACM-Reference-Format}
\bibliography{seis}

\end{document}